\documentclass[preprint,aps]{revtex4}

\usepackage{graphicx}
\usepackage{amssymb,amsmath}
\usepackage{caption}

\begin{document}

\title{Fermi Arcs vs. Fermi Pockets in Electron-doped Perovskite Iridates}
%
%
% The list of authors
%
\author{Junfeng He$^{1}$\footnote[1]{These authors contributed equally to this work.}\footnote[2]{Correspondence and requests for materials should be addressed to R.-H.H.(here@bc.edu); A.B.(ar.bansil@neu.edu) or J.H.(junfeng.he@bc.edu).}, H. Hafiz$^{2*}$, Thomas R. Mion$^{1}$, T. Hogan$^{1}$, C. Dhital$^{1,6}$, X. Chen$^{1}$, Qisen Lin$^{1}$, M. Hashimoto$^{3}$, D. H. Lu$^{3}$, Y. Zhang$^{4}$, R. S. Markiewicz$^{2}$, A. Bansil$^{2\dag}$, S. D. Wilson$^{1,5}$, and Rui-Hua He$^{1\dag}$}

\affiliation{
\\$^{1}$Department of Physics, Boston College, Chestnut Hill, MA 02467, USA
\\$^{2}$Physics Department, Northeastern University, Boston, MA 02115, USA
\\$^{3}$Stanford Synchrotron Radiation Lightsource, SLAC National Accelerator Laboratory, Menlo Park, CA 94025, USA
\\$^{4}$International Center for Quantum Materials, Peking University, Beijing 100871, China
\\$^{5}$Materials Department, University of California Santa Barbara, Santa Barbara, CA 93106, USA
\\$^{6}$Chemical and Engineering Materials Divison, Oak Ridge National Laboratory, Oak Ridge, TN 37831, USA}


\maketitle

\newpage
\begin{center}
\textbf{\large\centering{Abstract}} \\
\end{center}
We report on an angle resolved photoemission (ARPES) study of bulk electron-doped perovskite iridate, (Sr$_1$$_-$$_x$La$_x$)$_3$Ir$_2$O$_7$. Fermi surface pockets are observed with a total electron count in keeping with that expected from La substitution. Depending on the energy and polarization of the incident photons, these pockets show up in the form of disconnected ``Fermi arcs", reminiscent of those reported recently in surface electron-doped Sr$_2$IrO$_4$. Our observed spectral variation is consistent with the coexistence of an electronic supermodulation with structural distortion in the system.

\newpage
Many exotic phenomena take place after metallicity sets in when Mott insulators are doped with carriers. A spectacular example is provided by the $3d$ transition-metal copper oxides (cuprates) in which high-temperature superconductivity can be realized by doping the parent Mott state \cite{DopingMott}. Many aspects of the transition of a Mott insulator to a metal, however, remain poorly understood. In particular, the observation of ``Fermi arcs," which are disconnected gapless segments of an otherwise gapped Fermi surface in the normal or pseudogap state of the cuprates \cite{Yoshida,Marshall,Norman,KMShen,Kanigel,WSLee,Hossain,valla,MengJQ} presents a theoretical challenge. It is important, therefore, to establish the extent to which Fermi arcs are a universal feature of the electronic spectrum of a doped Mott insulator, signaling the emergence of fundamentally new physics in the material.

Perovskite strontium iridium oxides (iridates), the $5d$ electronic counterpart of the cuprates, have attracted much recent interest as a playground for understanding the effects of spin-orbit coupling in the presence of strong Coulomb interactions \cite{KimPRL,KimScience,Moon,VidyaNM}. Spin-orbit coupling can split the $t_{2g}$ manifold into $J=3/2$ and $J=1/2$ bands. In this way, the effective width of the valence bands is reduced, so that a moderate on-site Coulomb repulsion $U$ can now be sufficient to cause a further splitting of the $J=1/2$ band into upper and lower Hubbard bands. Such a ``spin-orbit Mott insulator", when doped, might serve as an ideal system to test the universality of various emergent phenomena found in cuprates and other doped Mott insulators. Notably, it has been proposed that superconductivity may be realized in the pervoskite iridates with electron doping \cite{FWang}. Remarkably, Fermi arcs have been reported recently in Sr$_2$IrO$_4$ based on surface electron doping via potassium deposition \cite{arc214}, raising many open questions, such as: Is there any conventional Fermi surface segment, which might coexist with Fermi arcs but may be masked due to unfavorable experimental conditions \cite{DLFengbilayer,YDbilayer1,YDbilayer2,arun1}? Is the existence of Fermi arcs a property of the electron-doped iridates independent of the doping method? Are the Fermi arcs a universal feature of the electron-doped pervoskite iridates irrespective of the material families as in the case of the hole-doped cuprates?

In order to address some of the aforementioned questions, we have carried out an ARPES study of perovskite iridate system, Sr$_3$Ir$_2$O$_7$. By substituting Sr atoms with La, electrons are doped into the bulk material, which leads to an insulator-metal transition at $x\sim0.033$ \cite{Cao3272}. Fermi pockets are observed with a total area consistent with the carrier concentration induced by La doping. The momentum location and overall shape of the observed pockets agree well with the corresponding first-principles calculations. In particular, we find that different types of low-energy electronic states in (Sr$_1$$_-$$_x$La$_x$)$_3$Ir$_2$O$_7$ ($x=0.066$) can be excited selectively by tuning the incident photons. Spectral intensity associated with parts of the Fermi pockets is suppressed due to matrix element effects when particular combinations of photon energy and polarization are used, giving these pockets the appearance of apparent open-ended Fermi arcs. The manner in which this partial spectral suppression occurs is consistent with the existence of an electronic supermodulation in the system. Our results, which are based on bulk doping, suggest that Fermi arcs are at least not universal to electron-doped perovskite iridates. Whether they exist in electron-doped Sr$_2$IrO$_4$ should be further scrutinized with the photoemission matrix element effects taken into account.

Single crystals of (Sr$_1$$_-$$_x$La$_x$)$_3$Ir$_2$O$_7$ ($x=0.066$) were grown by flux methods similar to the previous studies \cite{chetanPRB}. ARPES experiments were performed at Beamline 5-4 of the Stanford Synchrotron Radiation Laboratory with an energy resolution of $\sim$ 9 meV using a range of photon energies and polarizations. The Fermi level of the sample was referenced to that of a polycrystalline Au specimen in electrical contact with the sample. Measurements were done at 30 K with a base pressure of better than 3$\times$10$^{-11}$ torr.

Fermi pockets are observed in our sample based on our Fermi surface mapping shown in Fig. 1a, which is consistent with the metallic nature of the bulk material. In contrast to the cuprates, perovskite iridates crystallize in a nearly tetragonal structure with small rotations of the IrO$_6$ octahedra about the c-axis. The corresponding Brillouin zone (BZ) (gray dashed line in Fig. 1a) is smaller than that for the undistorted square lattice (white dashed line). In order to allow a direct comparison with the cuprates, however, we present our results in the undistorted (larger) BZ hereafter. Note that (Sr$_1$$_-$$_x$La$_x$)$_3$Ir$_2$O$_7$ contains two IrO$_2$ planes with two Ir sites in each undistorted unit cell, which is to be contrasted with the case of the single-layer Sr$_2$IrO$_4$. Incident light with $p$-polarized 25 eV photons (see Fig. 2a for the experimental setup and identification of $p$ polarization) was used for this particular measurement. Fermi pockets (red dashed ellipses are to guide the eye) can be seen clearly, whose number, shape and location in momentum agree with our first-principles calculations (Fig. 1b). In particular, considering eight Fermi pockets in the undistorted BZ yields an electron count of about 0.096 electrons/Ir or $x$$\sim$0.064 in (Sr$_1$$_-$$_x$La$_x$)$_3$Ir$_2$O$_7$, which is consistent with the La substitution level $x=0.066$ determined independently via energy-dispersive x-ray spectroscopy measurements.

In order to gain further insight, energy-momentum dispersion along Cut1 in Fig. 1a was measured. Raw ARPES spectra (energy distribution curves, EDCs, Fig. 1c), second-derivative EDC (Fig. 1d) and MDC (momentum distribution curves) images (Fig. 1e) along this cut all clearly reveal the presence of two parabolic electron-like bands, which yield two electron pockets at the Fermi level. Each parabolic band contains two branches, each producing one Fermi crossing (k$_F$), four in total for the two bands (k$_{F1}$$\sim$k$_{F4}$ in Fig. 1d). k$_{F2}$ and k$_{F3}$ are largely overlapping because the two pockets are located very close to each other (Fig. 1a). Similar measurements along Cut2 (Fig. 1a) are shown in Figs. 1f-h, where the two parabolic bands are separated to the two ends of the cut.

Very different results are found when the incident polarization and/or photon energy is changed, see Fig. 2a for the experimental setup. Fig. 2 shows measurements taken along Cut1 with $p$-polarized 25 eV light (b-d), $s$-polarized 20 eV light (e-g), and $s$-polarized 25 eV light (h-j). The Fermi surface measured by $p$-polarized 25 eV photons is reproduced from Fig. 1a on an expanded scale in Fig. 2b for comparison. With this incident light, at least three branches (marked \#1, \#3 \& \#4) out of the four related to the two parabolic bands are clearly visible in the raw (Fig. 2c) as well as the second-derivative MDC image plots (Fig. 2d). The weak branch (\#2) can also be identified when the raw EDCs in Figs. 1c \& 2c are carefully inspected. In sharp contrast, only one branch of each parabolic band can be observed with $s$-polarized 20 eV light, showing two nearly parallel ``bands" in Fig. 2f-g. These ``bands" are actually only two branches (\#1 \& \#3) of the two parabolic bands, which give rise to two apparent Fermi arcs at the Fermi level with each being approximately one half of the Fermi pocket (cf. Fig. 2b,e). A more striking observation is that with $s$-polarized 25 eV light only one branch (\#1) of one parabolic band remains visible and the other band is entirely absent in both image plots (Fig. 2i-j). In this case, only one Fermi arc is seen in the Fermi surface map (Fig. 2h) (see Supplementary Fig. S1 for results with other photon energies).

The preceding observation of an apparent Fermi arc accompanies a ``pseudogap"-like behavior in the spectral function measured with $s$-polarized 25 eV light. By tracing (to the extent possible) the local maxima of spectral weight at the Fermi level both in and away from the arc region, one can define a hypothetical ``underlying Fermi surface" connecting their momentum positions, e.g., points 1$-$6 (green crosses in Fig. 3a). Fig. 3b shows the corresponding symmetrized EDCs along this ``underlying Fermi surface". A single peak at the Fermi level is seen at points 1$-$4, while a spectral weight suppression is observed at points 5 \& 6. The overall situation is reminiscent of the hole-doped cuprates in which nodal electronic states produce a gapless Fermi arc, while states away from that region have their spectral weight (pseudo-)gapped out from the Fermi level \cite{Yoshida,Marshall,Norman,KMShen,Kanigel,WSLee,Hossain,valla,MengJQ}.

However, results of Fig. 2 show clearly that the apparent Fermi arc in our iridate sample is merely one part of a closed pocket with the photoemission intensity of other parts being suppressed due to the use of $s$-polarized 25 eV light. Excitation with $p$-polarized 25 eV light reveals a clear band gap spanning the Fermi level beyond the pocket regions as shown by the band dispersion maps along the two high-symmetry directions (Fig. 3e-f). Consistent with these observations, our first-principles calculations suggest that the conduction band gives rise to two well-separated electron pockets near the M point while the valence band is entirely gapped with the band top located at the X point (Fig. 3g). Here, the EDC measurements along the hypothetical ``underlying Fermi surface" therefore show an energy gap away from the arc region (e.g., point 6 in Fig. 3e). Because of the broad energy linewidth of the valence band, its non-vanishing spectral weight extends up to the Fermi level, which causes an impression that the gap is not a full gap but a pseudogap. Note that this gap along the hypothetical ``underlying Fermi surface" is not monotonic in that we expect a larger gap in the intermediate region between the M and X points than near the X point. This can be seen in the experimental results in Fig. 3e, and also in Fig. 3b where the EDC at point 5 shows a larger gap than at point 6.

(Sr$_1$$_-$$_x$La$_x$)$_3$Ir$_2$O$_7$ displays quasi-three-dimensional transport characteristics \cite{Cao3272}, which could indicate the presence of a significant $k_z$ dispersion and variations in band dispersion and Fermi surface probed with different photon energies. Nevertheless, our observation that the Fermi pocket and an apparent arc can both be seen at the same photon energy (25 eV) with different polarizations cannot be ascribed to this factor. Calculations have also been performed to investigate the $k_z$ dependence of the electronic structures. While some of the valence bands exhibit $k_z$ dispersion, the conduction band which forms the Fermi surface is found to display little $k_z$ dependence, and thus the associated Fermi surface area is essentially independent of $k_z$ (see Ref. \cite{neg} for details). A more sensible explanation lies in photoemission matrix element effects \cite{arun1,arun2,richard}, which can suppress the spectral intensity from parts of a Fermi surface and even render the related dispersion branches invisible for certain photon energies and/or polarizations. The ARPES matrix element encodes differences in the way initial states of various symmetries and orbital characters are excited by photons of different energies and polarizations. This occurs naturally in the case the $J=1/2$ band through the coupling between the two neighboring IrO$_2$ planes, which leads to the formation of bonding ($|k_{BB}>$) and anti-bonding bands ($|k_{AB}>$) of distinct orbital characters. It has been demonstrated in the familiar bilayer cuprate material, Bi2212, that such bonding and anti-bonding bands can be selectively highlighted in ARPES by using different photon energies and/or polarizations \cite{DLFengbilayer,YDbilayer1,YDbilayer2,arun1,arun2}.

Another mechanism is band folding around a certain wave vector $Q$, which would mix states of wave vectors $|k+Q>$ and $|k>$. For the iridates, the likely $Q$ would be $(\pi,\pi)$. This can be a consequence of IrO$_6$ octahedral rotations, which are known to exist in the undoped as well as the doped (Sr$_1$$_-$$_x$La$_x$)$_3$Ir$_2$O$_7$\cite{Cao3272}. In related systems, Sr$_2$RuO$_4$ and Sr$_2$RhO$_4$, a similar structure distortion can lead to pairs of $|k>$ and $|k+Q>$ bands of similar photoemission intensity \cite{Sr2RuO4,Sr2RhO4}, reminiscent of our observation in (Sr$_1$$_-$$_x$La$_x$)$_3$Ir$_2$O$_7$. Alternatively, the $(\pi,\pi)$ wave vector in (Sr$_1$$_-$$_x$La$_x$)$_3$Ir$_2$O$_7$ could be a consequence of local antiferromagnetic correlations. A $J=1/2$ antiferromagnetic order has been found at $x=0$ \cite{chetanPRB}. It is likely that this long-range order persists over shorter length scales in the metallic region as observed in other lightly-doped Mott insulators \cite{Armitagereview}.

Our first-principles calculations suggest that a structural distortion or antiferromagnetic correlations could in principle open a partial gap and induce formation of pockets at the Fermi level (Supplementary Figs. S2a-c). However, the preceding calculations require a fairly large value of spin-orbit coupling strength or $U$, and yield gap sizes and valence-band dispersions that are substantially different from experimental observations. On the other hand, if we include both a structural distortion and antiferromagnetic correlations in the calculation simultaneously, the values of $U$ or spin-orbit coupling strength required to open a partial gap and produce Fermi pockets of a similar size are smaller (compare Supplementary Fig. S2d with Figs. S2b and c), yielding a reasonable overall agreement with experiment (cf. Supplementary Fig. S2d vs. Fig. 3f). We thus conclude that Fermi pockets in (Sr$_1$$_-$$_x$La$_x$)$_3$Ir$_2$O$_7$ arise from a subtle interplay between structural distortion, antiferromagnetism, spin-orbit coupling, and electron correlation effects.

Despite the uncertainty regarding the precise origin of the ($\pi$,$\pi$) order, it is reasonable to designate the four branches of the two parabolic bands we have observed with $p$-polarized 25 eV light (\#1$\sim$\#4) as $|k_{BB}+Q>$, $|k_{AB}>$, $|k_{AB}+Q>$, and $|k_{BB}>$ states respectively(Fig. 2k), which all arise from the putative single $J=1/2$ band. It appears that $s$ polarization at both 20 eV and 25 eV suppresses photoemission from the $|k>$ states, while the 25 eV light preferentially excites the $|k_{BB}+Q>$ state (cf. Fig. 2e-j).

Our observation of Fermi arcs (Fig. 2h-j) looks superficially similar to that on Sr$_2$IrO$_4$ reported recently with surface electron doping \cite{arc214}. The shape and momentum location of the Fermi arcs seen in both cases are similar. However, we note two differences: The Fermi arcs of Sr$_2$IrO$_4$ show a nontrivial temperature dependence, and cross the antiferromagnetic (smaller) Brillouin zone boundary; effects not seen in Sr$_3$Ir$_2$O$_7$. It will be interesting to analyze details of spectral intensities and their dependencies on the energy and polarization of incident photons to gain further insight into the present spectra and their differences from the corresponding ARPES results on the Sr$_2$IrO$_4$ system. As far as Sr$_3$Ir$_2$O$_7$ is concerned, our study points instead to the existence of Fermi pockets, and leads to a fundamentally different picture for the low-lying excitations of the metallic perovskite iridates\cite{Torre}. Its clear departure from the Fermi-arc picture can be due to the possibility that the two materials have different properties, with Sr$_2$IrO$_4$ supporting Fermi arcs but Sr$_3$Ir$_2$O$_7$ harboring Fermi pockets with electron doping. This scenario would then imply that Fermi arcs are not a universal feature of electron-doped perovskite iridates, in sharp contrast to underdoped cuprates where Fermi arcs appear to be a generic feature of all single and bilayer families \cite{Yoshida,Marshall,Norman,KMShen,Kanigel,WSLee,Hossain,valla,MengJQ}. Another possibility is that Fermi arcs reported in Sr$_2$IrO$_4$ are tied to a unique surface state, which is only realized with surface potassium deposition and may not exist in bulk doped materials. These two possibilities could be distinguished via a comparative study of bulk- (La-doped) and surface-doped Sr$_2$IrO$_4$. Our study also points to keeping the importance of the ARPES matrix element in mind when adducing the presence of Fermi arcs vs. pockets from the spectra taken under particular experimental conditions.\\

\noindent{\textbf{Acknowledgement} } The work at Boston College was supported by a BC startup fund (J.H., R.-H.H.), the US NSF CAREER Awards DMR-1454926 (R.-H.H., in part) and DMR-1056625 (T.H., C.D., X.C., S.D.W.), and NSF Graduate Research Fellowship DGE-1258923 (T.R.M.). The work at NEU was supported by the DOE, BES grant number DE-FG02-07ER46352, and benefited from NEU's ASCC and the allocation of supercomputer time at NERSC through DOE grant DE-AC02-05CH11231. ARPES experiments were performed at the SSRL supported by the US DOE, BES Contract No. DE-AC02-76SF00515.\\

\noindent{\textbf{Competing Interests} } The authors declare that they have no competing financial interests.\\

\noindent{\textbf{Author Contributions} } J.H. proposed and designed the research with suggestions from R.-H.H.. J.H. and T.R.M. carried out the ARPES measurements with help from Q.L.. T.H. grew the samples and worked with C.D. and X.C. on sample characterizations. H.H. and R.S.M. performed the first-principles calculations and provided theoretical guidance. M.H. and D.H.L. maintained the ARPES beamline and endstation. J.H. analyzed the data with the help from Y.Z.. J.H. wrote the paper with key inputs from H.H., T.R.M., S.D.W., R.S.M., A.B. and R.-H.H.. R.-H.H., S.D.W., and A.B. are responsible for project direction, planning and infrastructure.\\

%%\vspace{3mm}

\newpage
\begin{figure*}[tbp]
\begin{flushleft}
\caption{\textbf{Observation of Fermi pockets in electron doped (Sr$_1$$_-$$_x$La$_x$)$_3$Ir$_2$O$_7$.}(a) Fermi surface consisting of 8 electron pockets in the first Brillouin zone (BZ) of the undistorted square lattice. Red dashed ellipses are drawn as guides to the eye. BZ boundary is marked by white dashed lines. The BZ for the distorted structure (gray dashed lines), which reflects IrO$_6$ octahedral rotations, is smaller. The undistorted BZ is useful for making a direct comparison with the cuprates. (b) Theoretical Fermi surface. (c) Photoemission spectra (energy distribution curves, EDCs) of the electron-like bands along cut1 in (a). Black triangles mark the parabolic dispersions, which are also evident in the corresponding second-derivative EDC (d) and MDC (e) images. (f-h) Similar to (c-e), respectively, but for cut2 in (a).}
\end{flushleft}
\end{figure*}

\newpage

\begin{figure*}[tbp]
\begin{flushleft}
\caption{\textbf{Measurements with different incident light.} (a) A schematic of experimental setup, where with $p$ polarization geometry, the electric field E lies in the pink plane that is defined by the incident light and lens axis of the electron analyzer, and E is perpendicular to this plane with $s$ polarization geometry. Here, $p$ and $s$ polarizations are defined loosely in that the angle between the pink plane and sample plane is $\sim$$75^{\circ}$ instead of $90^{\circ}$. This angle is determined by the momentum region we measure and the photon energy of the incident light when the analyzer is fixed. The incident angle of the light is $\sim$$45^{\circ}$. Fermi surface (b), photoemission intensity plot (c) along cut1, and the corresponding second-derivative MDC image (d) are first measured by $p$-polarized 25 eV light in which two Fermi pockets and parabolic electron-like bands can be well recognized. (e-g) and (h-j) are the same measurements but with $s$-polarized 20eV and 25eV incident light, respectively. Only parts of the two electron-like bands are visible (f-g and i-j), which results in apparent ``Fermi arc(s)" (e, h). (k) Schematic segments of Fermi pockets from four branches of the two electron-like bands.}
\end{flushleft}
\end{figure*}

\newpage
\begin{figure*}[tbp]
\begin{flushleft}
\caption{\textbf{Energy gaps along the ``underlying Fermi surface" and high symmetry dispersions.} (a) Symmetrized Fermi surface shown in a quadrant of the undistorted BZ measured by $s$-polarized 25eV incident light. (b) Symmetrized EDCs measured at points 1 to 6 with the same experimental conditions as in (a). The corresponding momentum positions are marked by green crosses in (a). A symmetrizing procedure is used to remove the effect of the Fermi function. (c) Calculated Fermi surface in a quadrant of the BZ. (d) The ``underlying Fermi surface" (dashed black line) is separated into three regions with the two regions away from the ``arc" being gapped. Second-derivative EDC images along cut3 (e) and cut2 (f) are measured by $p$-polarized 25eV incident light in order to show all the branches along the high symmetry directions. Cut3 is slightly curved with its center at the X point in the experimental setup used. The two ends of this cut are close to but not at the M points, which have captured a tail from the electron-like band near M. The pink arrow in (e) indicates the momentum location of point 6 in (a\&b). (g) First principles calculations. Spin-orbit coupling strength is set at 1.9 times the GGA + U self-consistently obtained value and U=0.727eV.}
\end{flushleft}
\end{figure*}

\newpage
\includegraphics[width=1.0\columnwidth,angle=0]{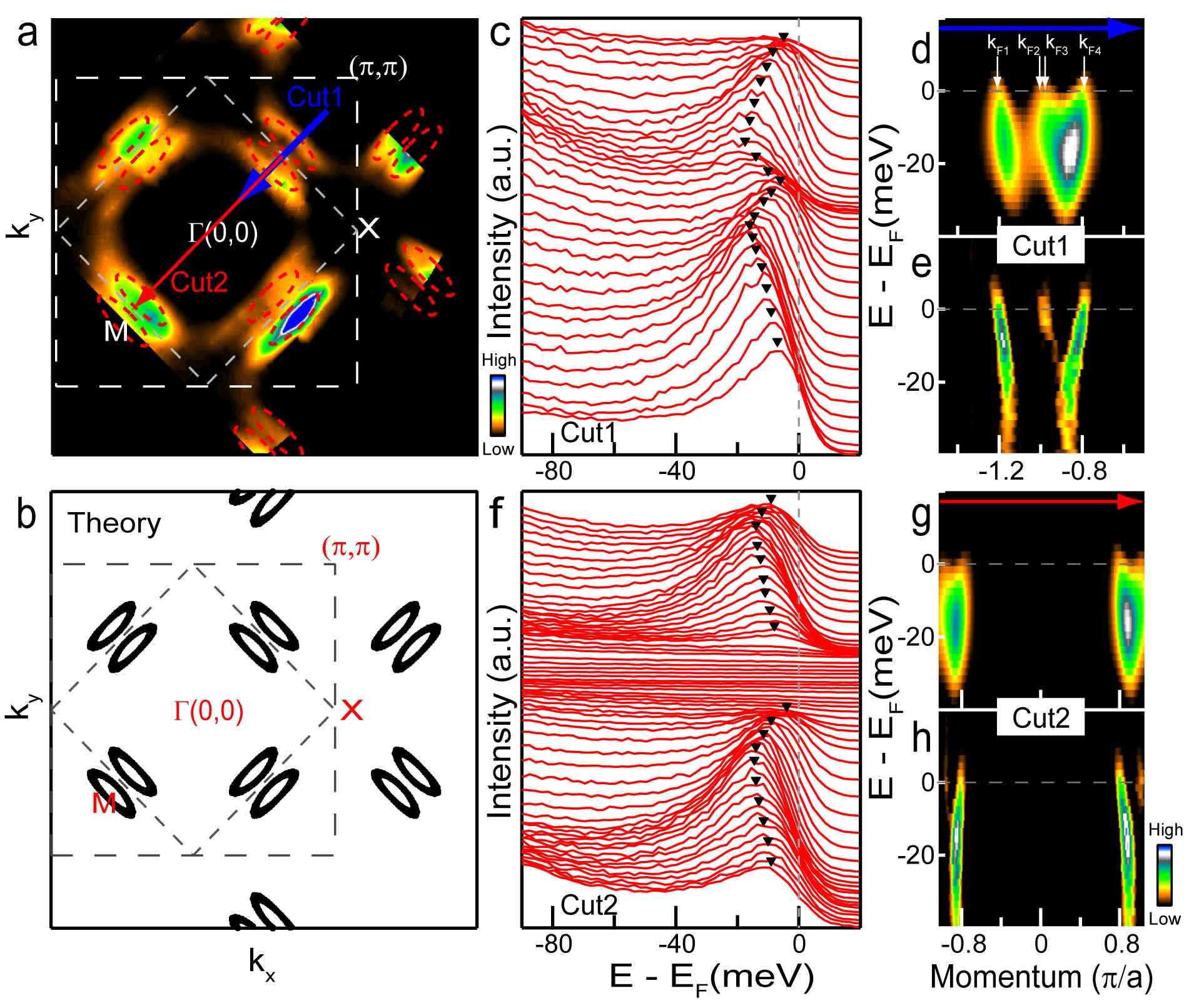}
\newpage
\includegraphics[width=1.0\columnwidth,angle=0]{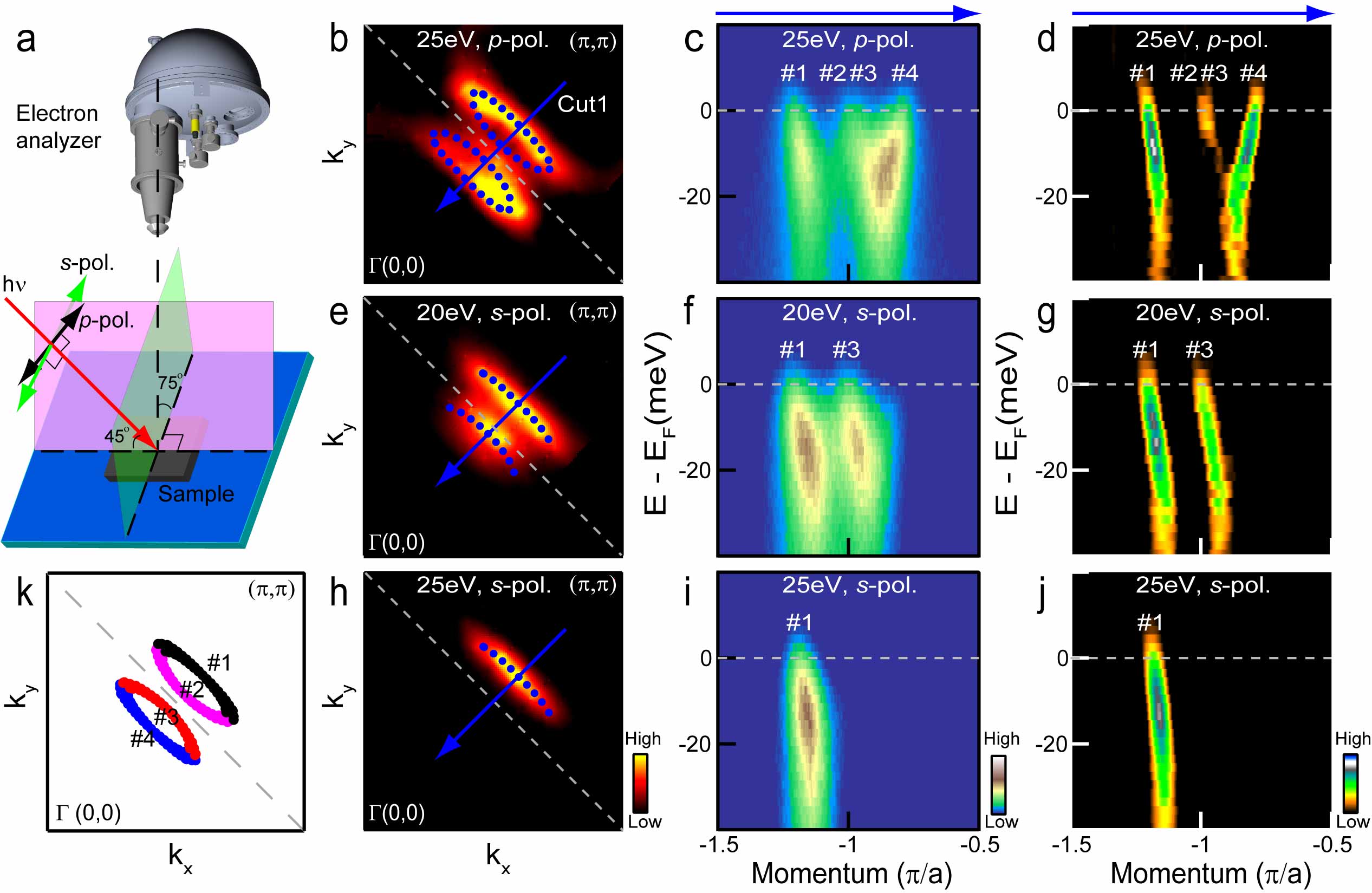}
\newpage
\includegraphics[width=1.0\columnwidth,angle=0]{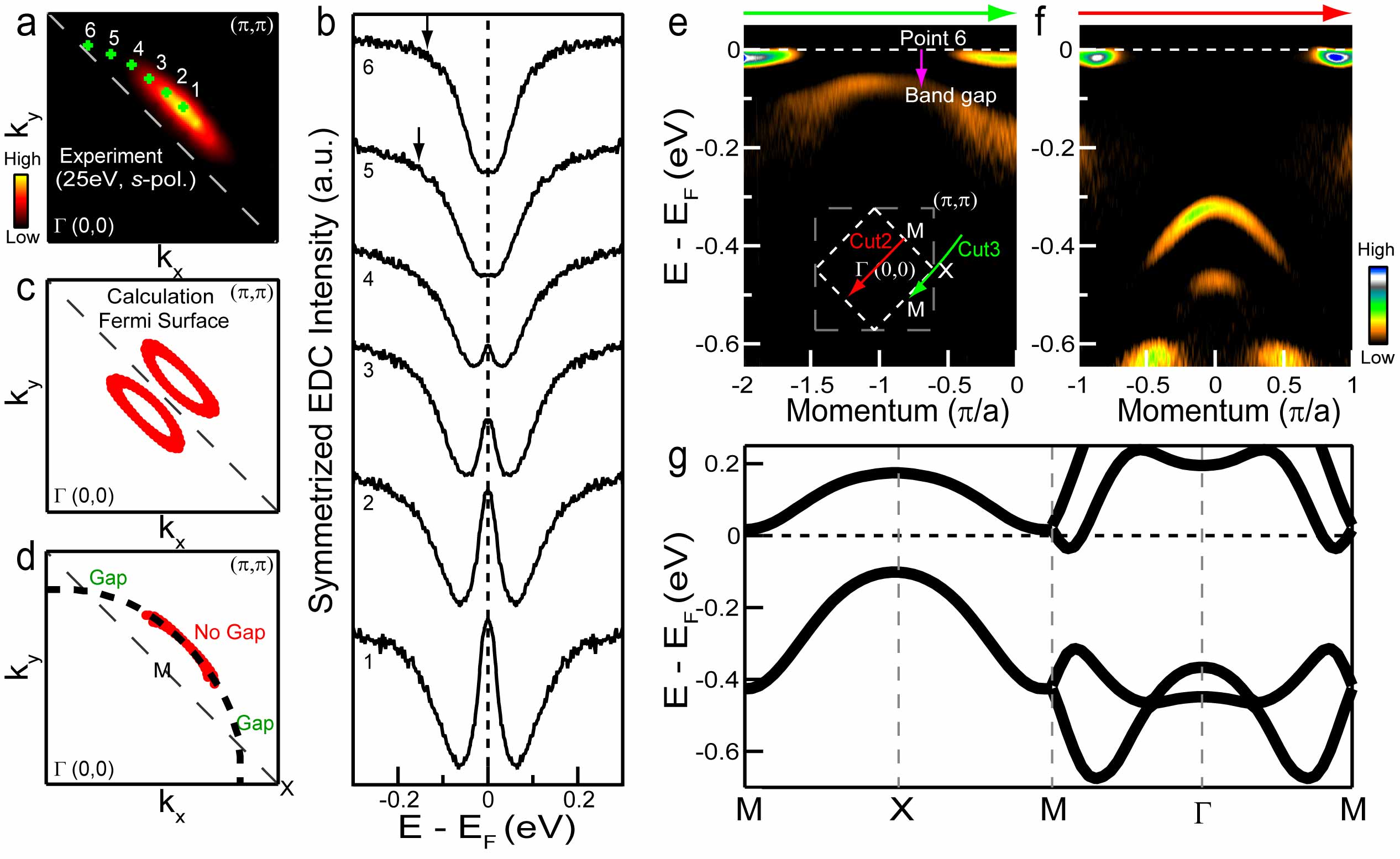}

\newpage
\begin{center}
\textbf{\large\centering{\emph{Supplementary Information for}}\\Fermi Arcs vs. Fermi Pockets in Electron-doped Perovskite Iridates}
\end{center}

\renewcommand\figurename{Fig. S}

\setcounter{figure}{0}
\renewcommand{\thefigure}{\arabic{figure}}

\noindent{\textbf{ARPES spectra along the (0,0)-($\pi$,$\pi$) cut at other photon energies and polarizations.} }\\

Measurements along the (0,0)-($\pi$,$\pi$) cut (the same momentum cut as that  shown in Fig. 2 of the main text) with $p$-polarized 23 eV and 30 eV light are shown in Fig. S1. It is clear that \#1 and \#2 bands are strongly suppressed with $p$-polarized 23 eV light (panels a and b of Fig. S1), while \#3 and \#4 bands are enhanced. On the other hand, the corresponding spectra at 30 eV in panels c and d of Fig. S1 preferentially enhance \#1 and \#2 bands. These results further highlight strong ARPES matrix element effects in this system.\\

\noindent{\textbf{Interplay of effects of structural distortion and antiferromagnetic correlations on the low-lying band structure.} }\\

Here we consider results of first-principles calculation without the presence of octahedral rotations (undistorted lattice). By considering antiferromagnetic correlations (AFM), Fermi pockets are seen to be clearly reproduced near M in panel a of Fig. S2. This is also seen in the related dispersion in panel b in which two electron-like bands appear near M, which produce the Fermi pockets. We have also carried out calculations in the paramagnetic state (PM) for the distorted lattice. Using a larger spin-orbit coupling strength, two similar electron-like bands and Fermi pockets can be produced near M (Fig. S2c). Note, however, that the calculations in panels a-c assume a fairly large value of spin-orbit coupling strength or $U$, and yield gap sizes and valence-band dispersions which differ substantially from the corresponding experimental results. On the other hand, if we include both a structural distortion and antiferromagnetic correlations in the calculation simultaneously, the values of $U$ or spin-orbit coupling strength required to open a partial gap and produce Fermi pockets of a similar size are smaller [compare panel (d) with panels (b) and (c)], yielding a reasonable overall agreement with experiment as seen by comparing panel (d) here with main Fig. 3f. We thus conclude that Fermi pockets in (Sr$_1$$_-$$_x$La$_x$)$_3$Ir$_2$O$_7$ arise from a subtle interplay between structural distortion, antiferromagnetism, spin-orbit coupling, and electron correlation effects.

\newpage
\begin{figure*}
\begin{center}
\includegraphics[width=1.0\columnwidth,angle=0]{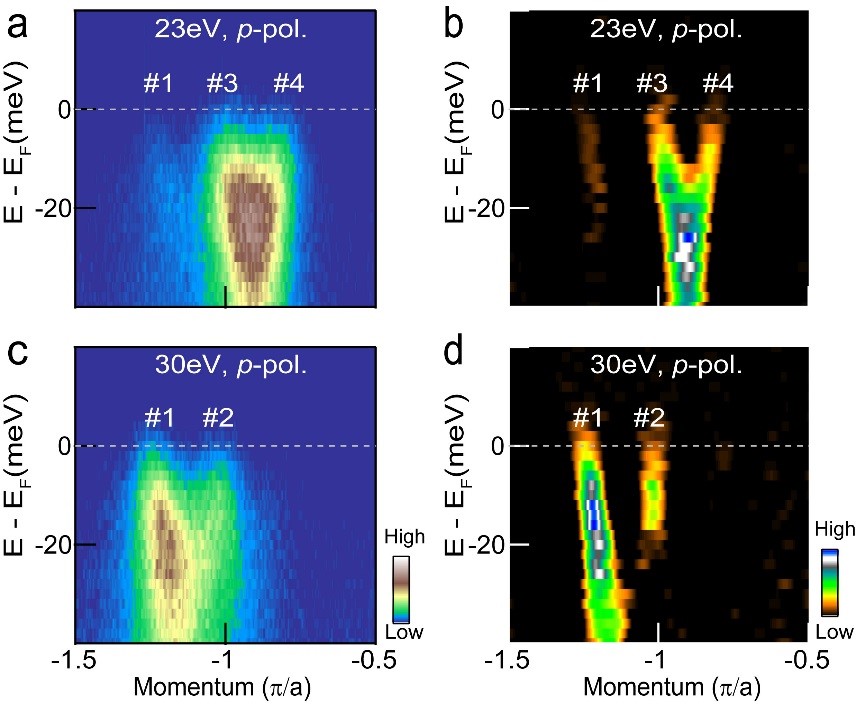}
\end{center}
\caption{\textbf{Measurements along the (0,0)-($\pi$,$\pi$) cut using other photon energies and polarizations.} Photoemission intensity plot (a) and the corresponding second-derivative MDC image (b) using $p$-polarized 23 eV light. (c \& d), the same as (a \& b) but with $p$-polarized 30 eV light. The momentum cut is the same as that shown in Fig. 2 of the main text.}
\end{figure*}

\newpage

\begin{figure*}
\begin{center}
\includegraphics[width=1.0\columnwidth,angle=0]{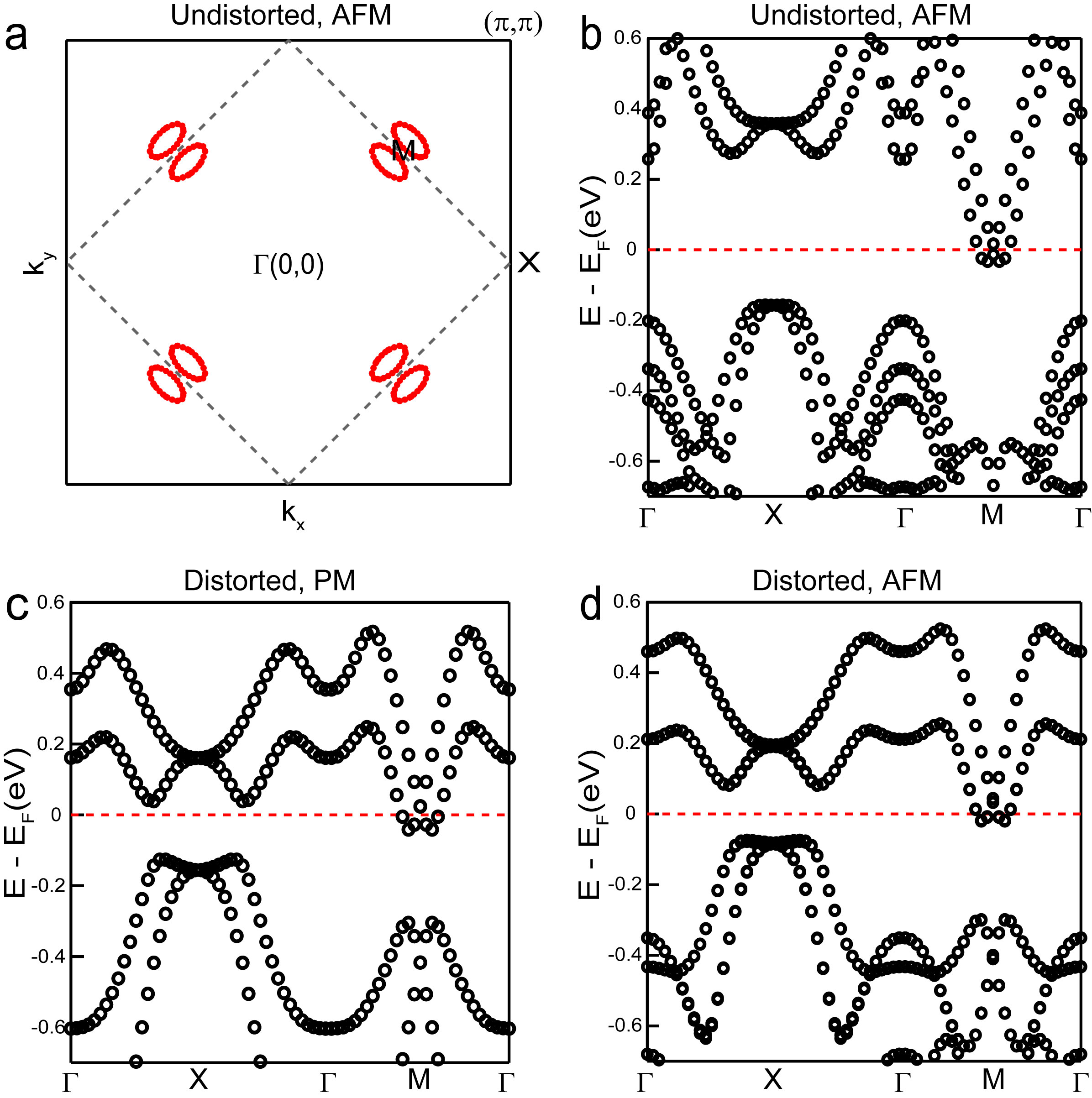}
\end{center}
\caption{\textbf{Interplay of effects of structural distortion and antiferromagnetic correlations on the low-lying band structure.} Fermi surface (a) and dispersion along high-symmetry directions in the Brillouin zone (b) for the undistorted crystal structure (i.e., without octahedral rotations) including antiferromagnetic correlations. $U=2$ eV and spin-orbit coupling strength is set at 1.9 times the GGA+U self-consistently obtained value, i.e. SOC=1.9. (c) Band structure for the paramagnetic state (PM) with distorted lattice. SOC=2.5. (d) Band structure when both the structural distortion and antiferromagnetic correlations ($U=0.727$ eV, SOC=1.9) are included in the computation.}

\end{figure*}

\end{document}